\documentclass[10pt,letterpaper]{article}

\usepackage{opex3}
 \usepackage[utf8x]{inputenc}
\usepackage{amssymb}
\usepackage{siunitx}
\usepackage[overload]{empheq}

\newcommand{\e}{\mathrm{e}}
\newcommand{\ic}{i}
\newcommand{\B}{\mathbf}
\newcommand{\tens}[1]{\smash{\underline{\underline{#1}}}}
\newcommand{\re}{\mathfrak{Re}}

\begin{document}

\title{Adaptive perfectly matched layer for Wood's anomalies in diffraction gratings}

 \author{Benjamin Vial,\textsuperscript{1,2,4,*} Fr\'ed\'eric Zolla,\textsuperscript{1,3}  Andr\'e Nicolet,\textsuperscript{1,3} Mireille Commandr\'e,\textsuperscript{1,4} and St\'ephane Tisserand\textsuperscript{2}}
 \address{\textsuperscript{1}Institut Fresnel, Domaine universitaire de Saint J\'er\^ome, 13397 Marseille cedex 20, France\\
 \textsuperscript{2}Silios Technologies, ZI Peynier-Rousset, rue Gaston Imbert Prolong\'ee, 13790 Peynier, France\\
\textsuperscript{3}Aix-Marseille Universit\'e, Jardin du Pharo, 58 boulevard Charles Livon, 13284 Marseille Cedex 07, France\\
\textsuperscript{4}\'Ecole Centrale Marseille, Technopôle de Château-Gombert, 38 rue Joliot-Curie, 13451 Marseille Cedex 20, France}
\email{benjamin.vial@fresnel.fr}

\date{\today}

 \begin{abstract}
We propose an Adaptive Perfectly Matched Layer (APML)
to be used in diffraction grating modeling. With a properly tailored co-ordinate stretching
 depending both on the incident field and on grating parameters,
the APML may efficiently absorb diffracted orders near grazing angles (the so-called
Wood's anomalies). The new design is implemented in a finite element method (FEM) scheme
and applied on a numerical example of a dielectric slit grating.
Its performances are compared with classical PML with constant stretching coefficient.
 \end{abstract}

\section{Introduction}

Since their introduction by Bérenger in \cite{Berenger1994185} for the time dependent Maxwell's equations, Perfectly Matched Layers (PMLs) have become a widely
 used technique in computational physics. The idea is to enclose the area of interest by surrounded layers
which are absorbing and perfectly reflectionless.
Subsequent formulations (\cite{Lassas2001739,CambridgeJournals:1202020}) allow to
 consider PMLs as a complex change of co-ordinates, which implies equivalent material properties (\cite{Nicolet2008,fpcf}).
A properly designed change of variables preserves the solution in the region of
 interest while introducing exponential decay at infinity. It is then obvious
to truncate the problem to a finite domain, set a convenient boundary condition on the boundary of this domain
 and apply the FEM. In a scattering problem, the ``traditional co-ordinate stretching''
 is frequency dependent, yielding a constant attenuation of propagating waves. However, these
 kinds of PMLs are inefficient for periodic problems when dealing with grazing angles of diffracted orders, i.e.\ when the frequency is near a Wood's anomaly (\cite{Wood,Rayleigh}), leading to spurious reflexions and thus numerical pollution of the results.
 An important question in designing absorbing layers is thus the choice of their parameters : the PML thickness and the absorption coefficient.
To this aim, adaptive formulation have been set up, most of them employing a posteriori error estimate (\cite{Chen2005, Bao2005, Schadle2007}).
 In this paper, we propose Adaptive PMLs (APMLs) with a suitable co-ordinate stretching,
depending both on incidence and grating parameters, capable of efficiently absorbing propagating waves with nearly grazing angles.
In the first section, we detail our FEM formulation and expose the problems arising when dealing with Wood's anomalies. Section 2 is dedicated to the
mathematical formulation used to determine PML parameters adapted to any diffraction orders. In the last section, we provide a numerical example of a dielectric slit grating showing the efficiency of our approach in comparison with classical PMLs.


\section{Setup of the problem and notations}

\subsection{Governing equations and description of the structure}
We denote by $\B x$, $\B y$ and $\B z$ the unit vectors of an orthogonal co-ordinate system $Oxyz$.
We deal with time-harmonic fields, so that the electric and magnetic fields are represented by complex
vector fields $\B E$ and $\B H$ with a time-dependence in $\mathrm{exp}(-i\omega t)$,
which will be discarded in the sequel. Moreover, we denote $k_0=\omega/c$.\\

\begin{figure}[h!]
\centering
\includegraphics[width=0.65\columnwidth]{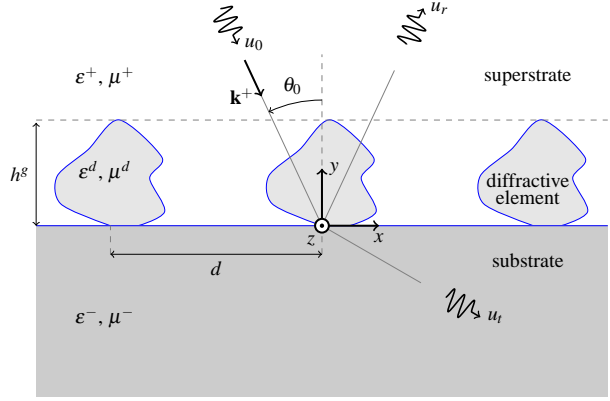}
\caption{Set up of the problem and notations}
\label{diffset}
\end{figure}

The materials in this paper may be $z$-anisotropic, so the tensor fields of relative permittivity $\tens{\varepsilon}$ and
 relative permeability $\tens{\mu}$ are of the following form :

\begin{equation}
\tens{\varepsilon}=
 \left(
  \begin{array}{ c c c}
\varepsilon_{xx} & \overline{\varepsilon_{a}} & 0 \\
 \varepsilon_{a} & \varepsilon_{yy} & 0 \\
  0 &  0 & \varepsilon_{zz}
  \end{array} \right)
\text{ and}\hspace{20pt}
\tens{\mu}=
 \left(
  \begin{array}{ c c c}
\mu_{xx} & \overline{\mu_{a}} & 0 \\
 \mu_{a} & \mu_{yy} & 0 \\
  0 &  0 & \mu_{zz}
  \end{array} \right),
\end{equation}

where the coefficients $\varepsilon_{xx}$, $\varepsilon_{aa}$,...$\mu_{zz}$ are (possibly) complex valued
 functions of $x$ and $y$, and where $\overline{\varepsilon_{a}}$ (resp. $\overline{\mu_{a}}$) is the complex conjugate of
$\varepsilon_{a}$ (resp. ${\mu_{a}}$).

The grating is illuminated by an incident plane wave of wave vector defined by the angle $\theta_0$ :
$\B k^+ =\alpha^+\B x+\beta^+\B y =k^+(\sin{\theta_0}\B x-\cos{\theta_0}\B y)$.
 Its electric (resp. magnetic) field is linearly polarized along the $z$-axis, this is the
 so-called transverse electric or s-polarization case (resp. transverse magnetic or p-polarization case) :
\begin{equation}
 \B E^0=\B A_e^0 \e^{(\ic\B k^+\cdot\B r)}\B z \hspace{20pt} (\text{resp. } \B H^0=\B A_m^0 \e^{(\ic\B k^+\cdot\B r) }\B z)
\end{equation}

where $\B A_e^0$ (resp. $\B A_m^0$) are arbitrary complex numbers and $\B r =(x,y)^{\rm T}$.

The diffraction problem we are dealing with is to solve Maxwell's equations in  harmonic regime, i.e.\ to
 find the unique solution $(\B E,\B H)$ of :

\begin{subequations}
\begin{align}[left=\empheqlbrace]
&\B{\mathrm{ curl}} \,\B E = \ic \omega \mu_0 \tens{\mu} \,\B H
\label{maxwell2DE}\\
&\B{\mathrm{ curl}} \,\B H = -\ic \omega \varepsilon_0 \tens{\varepsilon} \,\B E ,
\label{maxwell2DH}
\end{align}
\label{systmaxwell}
\end{subequations}

such that the diffracted field $\B E^{\rm d}:=\B E-\B E^{0}$ and $\B H^{\rm d}:=\B H-\B H^{0}$ satisfies an Outgoing Wave Condition (OWC, \cite{OWC}) and where $\B E$ and $\B H$ are quasi-periodic with respect to the $x$ co-ordinate.\\

Under the aforementioned assumptions, the diffraction problem in non conical
mounting can be separated in two fundamental scalar cases TE and TM.
Thus we search for a $z$-linearly polarized electric (resp.magnetic)
 field $\B E=e(x,y)\B z$ (resp. $\B H=h(x,y)\B z$). Denoting $\widetilde{\tens{\varepsilon}}$ and $\widetilde{\tens{\mu}}$
the $2\times2$ matrices extracted from $\tens{\varepsilon}$ and $\tens{\mu}$ :

\begin{equation}
 \widetilde{\tens{\varepsilon}}=
 \left(
  \begin{array}{ c c}
\varepsilon_{xx} & \overline{\varepsilon_{a}} \\
 \varepsilon_{a} & \varepsilon_{yy}
  \end{array} \right)
\hspace{10pt}\text{ and}\hspace{10pt}
 \widetilde{\tens{\mu}}=
 \left(
  \begin{array}{ c c}
\mu_{xx} & \overline{\mu_{a}}\\
 \mu_{a} & \mu_{yy}
  \end{array} \right),
\end{equation}

the functions $e$ and $h$ are solution of similar differential equations :

\begin{equation}
\mathcal{L}_{\tens{\xi},\chi}(u):=\mathrm{div}(\tens{\xi}\cdot \B{\mathrm{grad}}\; u)+k_0^2 \chi u=0,
\label{helm2D}
\end{equation}
with

\begin{equation}
 u=e, \hspace{5pt} \tens{\xi}=\tilde{\tens{\mu}}^{\rm T}/\mathrm{det}(\tilde{\tens{\mu}}),
\hspace{5pt} \chi=\varepsilon_{zz},
\end{equation}
for the TE case,
\begin{equation}
 u=h, \hspace{5pt} \tens{\xi}=\tilde{\tens{\varepsilon}}^{\rm T}/\mathrm{det}(\tilde{\tens{\varepsilon}}),
\hspace{5pt} \chi=\mu_{zz},
\end{equation}
for the TM case.\\

The gratings we are dealing with are made of three regions (see Fig. \ref{diffset}) :

\begin{itemize}
 \item \textit{The superstrate} ($y > h^g$) which is supposed to be
homogeneous, isotropic and lossless and characterized solely by its relative permittivity $\varepsilon^+$ and its relative
permeability $\mu^+$ and we denote $k^+ := k_0 \sqrt{\varepsilon^+\mu^+}$, where $k_0 := \omega/c$.
\item \textit{The substrate} ($y < 0$) which is supposed to be homogeneous and isotropic
and therefore characterized by its relative permittivity $\varepsilon^-$ and its relative permeability $\mu^-$ and we denote
$k^- := k_0 \sqrt{\varepsilon^-\mu^-}$
\item \textit{The groove region} ($0<y<h^g$), embedded in the superstrate,
 which can be heterogeneous, $z$-anisotropic and lossy, so that it is characterised by the tensor fields $\tens{\varepsilon}^g(x,y)$ and $\tens{\mu}^g(x,y)$.
The periodicity of the grooves along $Ox$ is denoted $d$.\\
\end{itemize}

\subsection{Implementation of the PMLs}
\label{impl_pml}

Transformation optics have recently unified various techniques in
computational electromagnetics such as the treatment of open problems, helicoidal geometries or the design of invisibility cloaks (\cite{Nicolet2008}).
These apparently different problems share the same concept of geometrical transformation, leading to equivalent material properties. A very simple and practical rule
can be set up (\cite{fpcf}) : when changing the co-ordinate system, all you have to do is to replace the initial materials properties $\tens{\varepsilon}$
 and $\tens{\mu}$ by equivalent material properties $\tens{\varepsilon}_s$ and $\tens{\mu}_s$ given by the following rule :

\begin{equation}
\tens{\varepsilon}_s=\B J^{-1}\tens{\varepsilon} \B J^{-\rm T}\mathrm{det}(\B J) \hspace{5pt}  \text{and~} \hspace{5pt} \tens{\mu}_s=\B J^{-1}\tens{\mu} \B J^{-\rm T}\mathrm{det}(\B J),
\label{equ_tranform}
\end{equation}
where $\B J$ is the Jacobian matrix of the co-ordinate transformation consisting of the partial derivatives of the new co-ordinates with respect to the original ones ($\B J^{-\rm T}$ is
the transposed of its inverse).\\
In this framework, the most natural way to define PMLs is to consider them as maps on a complex space $\mathbb{C}^3$, which co-ordinate change leads to equivalent permittivity and permeability tensors. We detail here the different co-ordinates used in this paper.
\begin{itemize}
 \item $(x,y,z)$ are the cartesian original co-ordinates.
\item $(x_s,y_s,z_s)$ are the complex stretched co-ordinates. A suitable subspace $\Gamma \subset \mathbb{C}^3$ is chosen (with three real dimensions)
such that $(x_s,y_s,z_s)$ are the complex valued co-ordinates of a point on $\Gamma$ (e.g.\ $x=\re(x_s)$, $y=\re(y_s)$, $z=\re(z_s)$).
\item $(x_c,y_c,z_c)$ are three real co-ordinates corresponding to a real valued parametrization of $\Gamma \subset \mathbb{C}^3$.
\end{itemize}

In this paper, we use rectangular PMLs (\cite{Zolla:07}) absorbing in the $y$-direction and
 we choose a diagonal matrix $\B J=\mathrm{diag}(1,s_y(y),1)$, where
 $s_y(y)$ is a complex-valued function defined by :

\begin{equation}
 y_s(y)=\int_0^y s_y(y')\rm d y'.
\label{ys}
\end{equation}
The expression of the equivalent permittivity and permeability tensors are thus :
\begin{equation}
\tens{\varepsilon}_s=
 \left(
  \begin{array}{ c c c}
s_y\varepsilon_{xx} & \overline{\varepsilon_a} & 0 \\
 \varepsilon_a & s_y^{-1}\varepsilon_{yy} & 0 \\
  0 &  0 & s_y\varepsilon_{zz}
  \end{array} \right)
 \hspace{10pt}  \text{and~} \hspace{10pt}
\tens{\mu}_s=
 \left(
  \begin{array}{ c c c}
s_y\mu_{xx} & \overline{\mu_a} & 0 \\
 \mu_a & s_y^{-1}\mu_{yy} & 0 \\
  0 &  0 & s_y\mu_{zz}
  \end{array} \right).
\label{tenspml}
\end{equation}

\begin{figure}[h!]
\centering
\includegraphics[width=0.5\columnwidth]{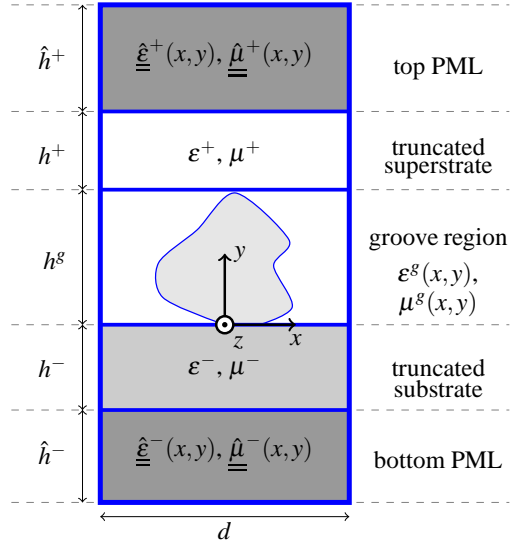}
\caption{The basic cell used for the FEM computation of the diffracted field $u_2^d$.}
\label{cell}
\end{figure}
Note that the equivalent medium has the same impedance than the original one as $\tens{\varepsilon}$ an $\tens{\mu}$ are transformed in the same way,
 which guarantees that the PML is perfectly reflectionless.\\
We are now in position to define the so-called substituted field $\B F_s=(\B E_s, \B H_s)$, solution of Eqs. (\ref{systmaxwell}) with $\tens{\xi}=\tens{\xi}_s$ and $\chi=\chi_s$.
$\B F_s$ equals the field $\B F$ in the region $y^b<y<y^t$ (with $y^b=-h^-$ and $y^t=h^g+h^+$), provided that $s_y(y)=1$ in this region (cf. Fig. \ref{cell}).
The complex co-ordinate mapping $y(y_c)$ is simply defined as the derivative of the stretching coefficient $s_y(y)$ with respect to $y_c$.
With simple stretching functions, we can obtain a reliable criterion on the decaying of the fields. A classical choice is :
\begin{equation}
 s_y(y)=
\begin{cases}
 \zeta^- & \mbox{if }y<y^b\\
1 & \mbox{if }y^b<y<y^t\\
 \zeta^+ & \mbox{if }y>y^t\\
\end{cases}
\end{equation}
where $\zeta^{\pm}=\zeta^{',\pm}+i\zeta^{'',\pm}$ are complex constants with $\zeta^{'',\pm}>0$.\\
In that case, the complex valued function $y(y_c)$ defined by Eq. (\ref{ys}) is explicitly given by :
\begin{equation}
 y(y_c)=
\begin{cases}
 y^b+\zeta^-(y_c-y^b)& \mbox{if }y_c<y^b\\
y_c & \mbox{if }y^b<y_c<y^t\\
 y^t+\zeta^+(y_c-y^t) & \mbox{if }y_c>y^t\\
\end{cases},
\end{equation}\\

Let's now study the behaviour of the field in the PMLs. In the substrate and the superstrate, according to Bloch's theorem, the diffracted field $u^d(x,y)$ can be written as :

\begin{equation}
u^d(x,y)=\sum_{n\in\mathbb{Z}}u_n^d(y)\e^{\ic\alpha_nx},
\label{rayl}
\end{equation}

where
\begin{equation}
 u_n^d(y)=\frac{1}{d}\int_{-d/2}^{d/2}u^d(x,y)\e^{-\ic\alpha_nx}\mathrm{d}x, \hspace{5pt} \mbox{with } \alpha_n=\alpha+\frac{2\pi}{d}n.
\end{equation}

Eventually, denoting
\begin{equation}
 {\beta_n^\pm}^2={k^\pm}^2-{\alpha_n^\pm}^2
\label{betan}
\end{equation}

 and accounting for an OWC, we obtain the expressions of the different
 diffraction orders :
\begin{equation}
 u_n^d(y)=
\begin{cases}
u_n^+(y)=r_n\e^{\ic\beta_n^+y} \mbox{ for } y>h^g\\
\\
u_n^-(y)=t_n\e^{-\ic\beta_n^-y} \mbox{ for } y<0.
\end{cases}
\end{equation}

We focus on the bottom PML, as similar considerations can be conducted for
 the top PML. We consider a diffraction order propagating in the substrate, which can be expressed
in the stretched co-ordinate system as :

\begin{equation*}
 u_{n,s}^{-}(y_c)=u_n^-(y(y_c))=t_n\e^{-\ic\beta_n^-[y^t+\zeta^-(y_c-y^t)]}.
\end{equation*}

The non oscillating part of this function is given by :

\begin{equation*}
U_n^{-}(y)=t_n\exp{\left((\beta_n^{',-}\zeta^{'',-}+\beta_n^{'',-}\zeta^{',-})y_c\right)},
\end{equation*}

 where $\beta_n^-=\beta_n^{',-}+\ic\beta_n^{'',-}$.
For a propagating order we have $\beta_n^{',-}>0$ and $\beta_n^{'',-}=0$, while for an evanescent order $\beta_n^{',-}=0$ and $\beta_n^{'',-}>0$.
 It is thus sufficient to take $\zeta^{',-}>0$ and $\zeta^{'',-}>0$ to ensure the exponential decay to zero of the
field inside the PML if it was of infinite extent.
 But, of course, for practical purposes, the height
 of the PML is finite and have to be suitably chosen. For this, two pitfalls must be avoided :
\begin{enumerate}
 \item The PML is too small compared to the skin depth. As a consequence, the electromagnetic wave cannot
 be considered as vanishing : this limited  PML is no longer reflectionless.
\item The PML is much larger than the skin depth.  In that case, a significant part of the PML is not useful,
 which gives rise to the resolution of  linear systems of large dimensions.
\end{enumerate}
It then remains to derive the skin depth, $l_n^-$, associated with the propagating order $n$.
This characteristic depth is defined as the length for which the field is divided by $e$ :
 \begin{equation*}
  U_n^{-}(y-l_n^-)=\frac{U_n^{-}(y)}{\e}.
 \end{equation*}

So $l_n^-=(\beta_n^{',-}\zeta^{'',-}+\beta_n^{'',-}\zeta^{',-})^{-1}$, and we take the larger value among the $l_n^-$ :
 \begin{equation*}
  l^-=\underset{n\in\mathbb{Z}}{\mathrm{max}}\,l_n^-.
 \end{equation*}

We set the height of the bottom PML region $\hat{h}^-=10l^-$.\\

\subsection{The FEM formulation}
Our FEM formulation is that described in \cite{Demesy:07,Demesy:09}. It relies on the fact that
the diffraction problem can be rigorously treated as an equivalent radiation problem with sources inside the diffractive object.
The expression of the source terms depends on the field $u_1$ solution of the annex problem of a simple interface,
and is known in closed form. In this context, the role of the PMLs is to damp the field radiated by the sources.
We set quasi-periodicity conditions on lateral boundaries and Neumann homogeneous conditions
 are imposed on the outward boundary of each PML. By so doing,
 the electromagnetic wave is not forced to vanish and the values of the computed field on these
boundaries give valuable information on
 the absorbing efficiency of the PMLs. The cell defined in Fig. \ref{cell} is meshed using $2$\textsuperscript{nd} order Lagrange elements (\cite{Ern}). In the numerical
examples in the sequel, the maximum element size is set to $\lambda_0/ (N_m\sqrt{\mathfrak{Re}(\varepsilon_r)})$, where $N_m$
 is an integer (between 6 and 10 is usually a good choice). The final algebraic system is solved using a direct solver (PARDISO).\\
Note that we use this FEM formulation in this paper but the results of the sequel about PMLs are general
and can be applied to any numerical method used for the modelling of diffraction gratings.

\subsection{Weakness of the classical PML for grazing diffracted angles}
\label{example_pb_classical_PML}
\begin{figure}[htb!]
\centering
\includegraphics[width=0.8\columnwidth]{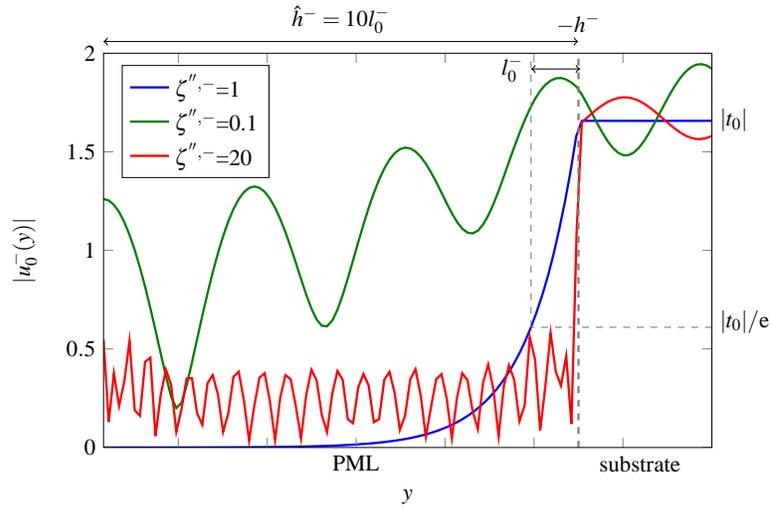}
\caption{Zero\textsuperscript{th} transmitted order by a grating with a rectangular cross section (see parameters in text, part \ref{example_pb_classical_PML}) for different values of
$\zeta^{'',-}$ : blue line, $\zeta^{'',-}=1$, correct damping; green line, $\zeta^{'',-}=0.1$, underdamping; red line, $\zeta^{'',-}=20$, overdamping.}
\label{behaviour_classical_PML}
\end{figure}
Resuming part \ref{impl_pml}, we consider only the bottom PML, as analogous conclusions holds for the top PML.
The efficiency of the classical PML fails for grazing diffracted angles, in other words when a given order appears/vanishes :
this is the so-called Wood's anomaly, well known in the theory of gratings. In mathematical words, there exists $n_0$ such that
$\beta_{n_0}^-\simeq 0$. The skin depth of the PML then becomes very large. To compensate it, one should increase the
 value of $\zeta^{'',-}$, but this gives rise to spurious numerical reflections due to an overdamping. For a fixed value of $\hat{h}^-$,
if $\zeta^{'',-}$ is too weak, the absorption in the PMLs is insufficient and the wave reflects on the outward boundary of the PML.
To illustrate these typical behaviours  (cf. Fig. \ref{behaviour_classical_PML}), we compute the diffracted field by a grating with a rectangular cross section of height $h^g=1.5$\si{\micro\meter}~and
width $L^g=3$\si{\micro\meter}~with $\varepsilon^g=11.7$, deposited on a substrate with permittivity $\varepsilon^-=2.25$. The structure is illuminated
 by a \textit{p}-polarized plane wave of wavelength $\lambda_0=10$\si{\micro\meter}~and of angle of incidence $\theta_0=10$\si{\degree}~in the air ($\varepsilon^+=1$). All materials are non magnetic ($\mu_r=1$) and the periodicity of the grating is $d=4$\si{\micro\meter}. We set $\hat{h}^-=10l_0^-$ and
$\zeta^{',-}=1$.

\section{Construction of an adaptive PML}
\label{part_stretch}
To overcome the problems pointed out in the preceding section, we propose a
co-ordinate stretching that rigorously treats the problem of Wood's anomalies.
The wavelengths ``seen'' by the system are very different depending on the order at stake :
\begin{itemize}
 \item if the diffracted angle $\theta_n$ is zero, the apparent wavelength $\lambda/\cos\theta_n$ is simply the incident wavelength,
\item if the diffracted angle is near $\pm \pi/2$ (grazing angle), the apparent wavelength $\lambda/\cos\theta_n$ is very large.
\end{itemize}
Thus if a classical PML is adapted to one diffracted order, it will not be for another, and
 vice versa. The idea behind the APML is to deal with each and every order when progressing in the absorbing medium.\\

Once again the development will be conducted only for the PML adapted to the substrate.
 We consider a real-valued co-ordinate mapping $y_d(y)$, the final
 complex-valued mapping is then $y_c(y)=\zeta^-y_d(y)$, with the complex constant
$\zeta^-$, with $\zeta^{',-}>0$ and $\zeta^{'',-}>0$, accounting for the damping of the PML medium.
We start by transforming Eq. (\ref{betan}), $n\mapsto\beta_n^-$ a function with integer argument into a function with real argument
continuously interpolated between the imposed integer values. Indeed, the geometric transformations associated to the PML has to
be continuous and differentiable in order to compute the Jacobian. For this purpose, we choose the parametrization :

\begin{equation}
 \alpha(y_d)=\alpha_0+\frac{2\pi}{d}\frac{y_d}{\lambda_0},
\end{equation}

so that the application $\beta^-$ defined
by ${\beta^-(y_d)}^2=k_0^2\varepsilon^--{\alpha(y_d)}^2$ is continuous.
Thus, the propagation constant of the $n^{\text{th}}$ transmitted order is given by
 $\beta_n^-=\beta^-(n\lambda_0)$. The key idea is to combine the complex stretching with
 a real non uniform contraction (given by the continuous function $y(y_d)$, Eq.(\ref{yyd})). This contraction is chosen in such a way that for each order
 $n$ there is a depth $y_d^n$ such that around this depth the apparent wavelength corresponding to the order in play is contracted to a value close to $\lambda_0$. 
 At that point of the PML, this order is perfectly absorbed thanks to the complex stretch.
 We thus eliminate first the orders with quasi normal diffracted angles at lowest depths up to grazing orders (near Wood's anomalies) 
 which are absorbed at greater depths. In mathematical words, the translation of previous considerations on the real contraction can be expressed as :
\begin{equation}
\exp{[-\ic\beta^-(y_d)y(y_d)]}=\exp{(-\ic k_0 y_d)}
\end{equation}
The contraction $y(y_d)$ is thus given by :
\begin{equation}
 y(y_d)=\frac{k_0y_d}{\beta^-(y_d)}=\frac{y_d}{\sqrt{\varepsilon^--(\sin{\theta_0}+y_d/d)^2}}
\label{yyd}
\end{equation}

The function $y(y_d)$ has two poles, denoted
$y_{d,\pm}^\star=d(\pm\sqrt{\varepsilon^-}-\sin{\theta_0})$. When $y_{d,\pm}^\star=\pm n\lambda_0$ with $n\in\mathbb{N}^\star$,
 $\beta^-(y_{d,\pm}^\star)=\beta^-(\pm n\lambda_0)=\beta_{\pm}^-=0$, i.e.\ we are on a Wood's anomaly associated
with the appearance/disappearance of the $\pm n^{\text{th}}$ transmitted order. We now search for the nearest
 point to $y_{d,\pm}^*$ associated
 with a Wood's anomaly, denoting :

\begin{equation*}
\begin{cases}
 n_+^\star/ \quad D_+=\underset{n_+^\star\in\mathbb{N}^\star}{\mathrm{min}}\,|y_{d,+}^\star-n_+^\star\lambda_0|\\
\\
 n_-^\star/ \quad D_-=\underset{n_-^\star\in\mathbb{N}^\star}{\mathrm{min}}\,|y_{d,-}^\star+n_-^\star\lambda_0| ,
\end{cases}
\end{equation*}
In a second step, we seek the point $y_d^0=n^\star\lambda_0$ such that :

\begin{equation}
 n^\star/ \quad D=\underset{n^\star\in\{n_+^\star,n_-^\star \}}{\mathrm{min}}\,(D^+,D^-),
\end{equation}

To avoid the singular behaviour at $y_d=y_{d,\pm}^\star$,
we continue the graph of the function $y_d(y)$ by a straight line tangent at $y_d^0$, which equation is $t_0(y_d)=s(y_d^0)(y_d-y_d^0)+y(y_d^0)$,
where $s(y_d)=\frac{\partial y}{\partial y_d}(y_d)$ is the so-called stretching coefficient.
The final change of co-ordinate is then given by~:
\begin{equation}
 \tilde{y}(y_d)=
\begin{cases}
y(y_d) \mbox{ for } y_d\leq y_d^0\\
\\
t_0(y_d) \mbox{ for } y_d>y_d^0.
\end{cases}
\end{equation}

\begin{figure}[h]
\centering
\includegraphics[width=0.6\columnwidth]{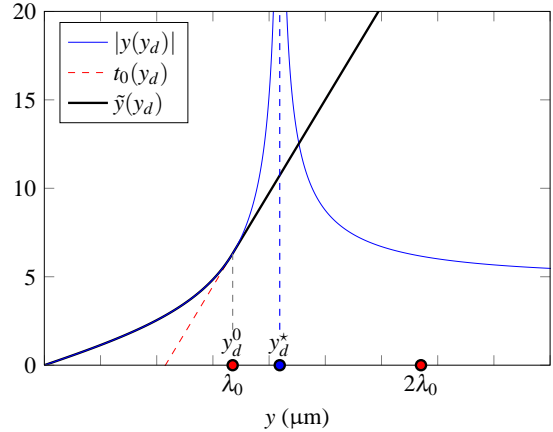}
\caption{Example of a co-ordinate mapping $\tilde{y}(y_d)$ used for the APML (black solid line). The graph of $y_d(y)$ (blue solid line) is continued
 by a straight line $t_0(y_d)$ tangent at $y_d^0$ (red dashed line) to avoid the singular behaviour at $y_d=y_{d}^\star$.}
\label{coef_stretch}
\end{figure}

Figure \ref{coef_stretch} shows an example of this co-ordinate mapping.

Eventually, the complex stretch $s_y$ used in Eq. (\ref{tenspml}) is given by :
\begin{equation}
 s_y(y_d)=\zeta^-\frac{\partial \tilde{y}}{\partial y_d}(y_d).
\end{equation}

Equipped with this mathematical formulation, we can tailor a layer that is doubly
perfectly matched :
\begin{itemize}
 \item to a given medium, which is the aim of the PML technique, through Eq. (\ref{equ_tranform}),
\item to all diffraction orders, through the stretching coefficient $s_y$, which
depends on the characteristics of the incident wave and on opto-geometric parameters
 of the grating.
\end{itemize}

\section{Numerical example}

We now apply the method described in the preceding parts to design an adapted bottom PML for the same example
 as in part \ref{example_pb_classical_PML}. The parameters are the same, and we choose the wavelength
of the incident plane wave close to the Wood's anomaly related to the $+1$ transmitted order
($\lambda_0=0.999 y_{d,+}^\star$).Moreover, we set the length of the PML $\hat{h}^-=1.1y_{d,+}^\star$ and choose absorption coefficients
$\zeta^{+}=\zeta^{-}=1+\ic$. For both cases (PML and APML), parameters are alike, the only difference being the complex stretch $s_y$.

\begin{figure}[h!]
\centering
\includegraphics[width=0.8\columnwidth]{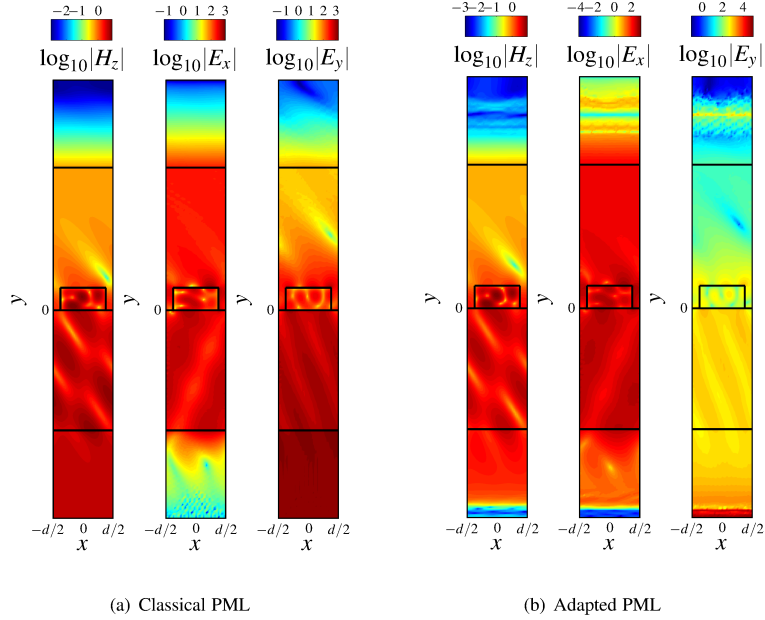}
\caption{Field maps of the logarithm of the norm of $H_z$, $E_x$ and $E_y$ for the dielectric slit grating at $\lambda_0=0.999 y_{d,+}^\star$ 
(same parameters as in part \ref{example_pb_classical_PML}). (a) : classical PML with inefficient damping of $H_z$ in the bottom PML. 
(b) : APML where the $H_z$ field is correctly damped in the bottom PML. For both cases the thickness of the PML is $\hat{h}^-=1.1y_{d,+}^\star$.}
\label{comp_PMLs}
\end{figure}

\begin{figure}[h!]
\centering
\includegraphics[width=0.7\columnwidth]{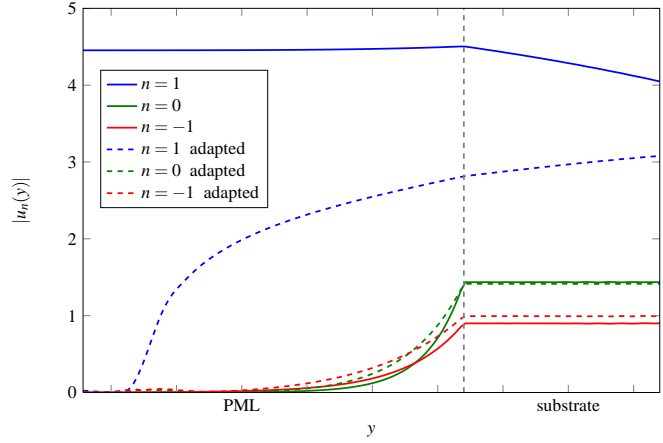}
\caption{Modulus of the $u_n$ for the three propagating orders with adapted (dashed lines) and classical PMLs (solid lines).
Note that the classical PMLs are efficient for all orders except for the grazing one ($n=1$) as expected. This drawback is bypassed when using the adaptive PML. }
\label{behaviour_adapted_PML}
\end{figure}

 The field maps of the norm of $H_z$, $E_x$ and $E_y$ are plotted in logarithmic scale on Fig. \ref{comp_PMLs}, for the case of a classical PML  and our APML.
We can observe that the field $H_z$ that is effectively computed is clearly damped in the bottom APML (leftmost on Fig. \ref{comp_PMLs}(b))
 whereas it is not in the standard case (leftmost on Fig. \ref{comp_PMLs}(a)), causing spurious reflections on the outer boundary.
The fields $E_x$ and $E_y$ are deduced from $H_z$ thanks to Maxwell's equations. The high values
of $E_y$ at the tip of the APML (rightmost on Fig. \ref{comp_PMLs}(b)) are due to very high values of the optical
equivalent properties of the APML medium (due to high values of $s_y$), which does not affect the accuracy of the computed field within the domain of interest.\\
Another feature of our approach is that it efficiently absorbs the grazing diffraction order, as illustrated on Fig. \ref{behaviour_adapted_PML} : the $+1$ transmitted order does not decrease in the standard
PML (blue solid line), and reaches a high value at $y=-\hat{h}^-$, whereas the same order
tends to zero as $y\rightarrow-\hat{h}^-$ in the case of the adapted PML (blue dashed line).\\
To further validate the accuracy of the method, we compare the diffraction efficiencies computed by our FEM formulation with PML and APML to those obtained by another method. We choose the Rigorous Coupled Wave Analysis (RCWA), also known as the Fourier Modal Method (FMM, \cite{Li}). For the chosen parameters, only the $0$\textsuperscript{th} order is propagative in reflexion and the orders $-1$, $0$ and $+1$ are non evanescent in transmission.
We can also check the energy balance $B=R_0+T_{-1}+T_{0}+T_{+1}$ since there is no lossy medium in our example.
Results are reported in Table \ref{comp_PML_RCWA}, and show a good agreement of the FEM with APML with the results from RCWA. 
On the contrary, if classical PML are used, the diffraction efficiencies are less accurate compared to those computed with RCWA. 
Checking the energy balance leads the same conclusions : the numerical result is perturbed by the reflection of the waves at the end of the PML if it is not
adapted to the situation of nearly grazing diffracted orders.\\

\begin{table}[h!]
 \centering
\begin{tabular}{|c|c|c|c|c|c|}
  \hline
   & $R_0$ & $T_{-1}$ & $T_{0}$ & $T_{+1}$ & $B$ \\
  \hline
  RCWA & 0.1570 & 0.3966 &  0.1783 & 0.2680 & 0.9999 \\
  FEM + APML & 0.1561 & 0.3959 & 0.1776 & 0.2703 & 0.9999 \\
  FEM + PML & 0.1904 & 0.4118 & 0.1927 & 0.2481 & 1.0430 \\
  \hline
\end{tabular}
\caption{Diffraction efficiencies $R_0$, $T_{-1}$, $T_{0}$ and $T_{+1}$ of the four propagating orders, and energy balance $B=R_0+T_{-1}+T_{0}+T_{+1}$, computed by three methods : RCWA (line 1), FEM formulation with APML (line 2), FEM formulation with classical PML (line 3).}
\label{comp_PML_RCWA}
\end{table}

\begin{figure}[h!]
\centering
\includegraphics[width=0.6\columnwidth]{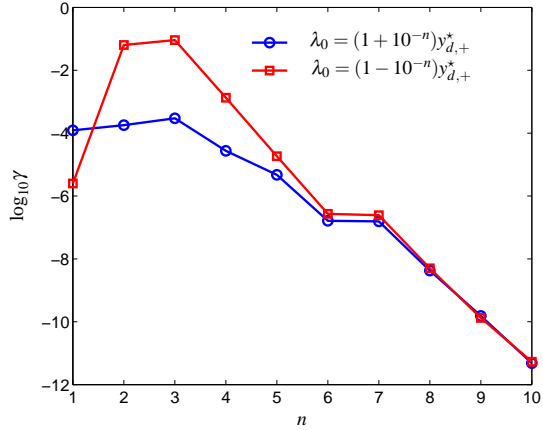}
\caption{Mean value of the norm of $H_z$
along the outer boundary of the bottom PML $\gamma=\langle|H_z(-\hat{h}^-)|\rangle_x$, for $\lambda_0$ approaching the
Wood's anomaly $y_{d,+}^\star$ by inferior values ($\lambda_0=(1-10^{-n})y_{d,+}^\star$, red squares)
 and by superior value ($\lambda_0=(1+10^{-n})y_{d,+}^\star$, blue circles) as a function of $n$.}
\label{comp_PML_lambda}
\end{figure}

Eventually, to illustrate the behaviour of the adaptive PML when the incident wavelength gets
 closer to a given Wood's anomaly, we computed the mean value of the norm of $H_z$
along the outer boundary of the bottom PML $\gamma=\langle|H_z(-\hat{h}^-)|\rangle_x$, when $\lambda_0=(1+10^{-n})y_{d,+}^\star$
and $\lambda_0=(1-10^{-n})y_{d,+}^\star$, for $n=1,2,...10$. The results are shown in
Fig. \ref{comp_PML_lambda}. As the wavelength gets closer to $y_{d,+}^\star$,
$\gamma$ first increases but for $n>3$, it decreases exponentially. However, in all cases, the value
of $\gamma$ remains small enough to ensure the efficiency of the PMLs.\\


\section{Conclusion}

In this paper, we have proposed an adaptive PML that can treat rigorously Wood's anomalies
 in numerical analysis of diffraction gratings. It is based on a complex-valued co-ordinate stretching that deals with
grazing diffracted orders, yielding an efficient absorption of the field inside the PML.
We provided an example in the TM polarization case (but similar results hold for the TE case),
 illustrating the efficiency of our method. The value of the magnetic field on the
outward boundary of the PML remains small enough to consider there is no spurious reflection.
The formulation is used with the FEM but can be applied to others numerical methods. Moreover, the generalization to the vectorial
three-dimensional case is straightforward : the recipes given in part \ref{part_stretch} do work irrespective of the dimension and whether
the problem is vectorial. In addition, although the designed co-ordinate stretching is specific to the context of Wood's anomalies, one can adapt this kind of non uniform PML to others wave problems by following the methodology exposed here.


\begin{thebibliography}{10}
\newcommand{\enquote}[1]{``#1''}

\bibitem{Berenger1994185}
J.-P. B\' erenger, \enquote{A perfectly matched layer for the absorption of
  electromagnetic waves,} Journal of Computational Physics \textbf{114}, 185 --
  200 (1994).

\bibitem{Lassas2001739}
M.~Lassas, J.~Liukkonen, and E.~Somersalo, \enquote{Complex riemannian metric
  and absorbing boundary conditions,} J. Math. Pure Appl. \textbf{80}, 739 -- 768 (2001).

\bibitem{CambridgeJournals:1202020}
M.~Lassas and E.~Somersalo, \enquote{{Analysis of the PML equations in general
  convex geometry},} P. Roy. Soc. Edinb. A \textbf{131}, 1183--1207 (2001).

\bibitem{Nicolet2008}
A.~Nicolet, F.~Zolla, Y.~Ould Agha, and S.~Guenneau, \enquote{{Geometrical
  transformations and equivalent materials in computational electromagnetism},}
  Compel \textbf{27}, 806--819 (2008).

\bibitem{fpcf}
F.~Zolla, G.~Renversez, A.~Nicolet, B.~Kuhlmey, S.~Guenneau, D.~Felbacq,
  A.~Argyros, and S.~Leon-Saval, \emph{Foundations of photonic crystal fibres}
  (Imperial College Press, London, 2012), 2nd ed.

\bibitem{Wood}
R.~Wood., \enquote{On a remarkable case of uneven distribution of light in a
  diffraction grating spectrum,} P. Phys. Soc. Lond. \textbf{18},
 269--275 (1902).

\bibitem{Rayleigh}
Lord~Rayleigh, \enquote{{Note on the remarkable case of diffraction spectra
  described by prof. Wood},} Philos. Mag. \textbf{14}, 60--65 (1907).

\bibitem{Chen2005}
Z.~Chen and X.~Liu, \enquote{An adaptive perfectly matched layer technique for
  time-harmonic scattering problems,} SIAM J. Numer. Anal. \textbf{43},
  645--671 (2005).

\bibitem{Bao2005}
G.~Bao, Z.~Chen, and H.~Wu, \enquote{Adaptive finite-element method for
  diffraction gratings,} J. Opt. Soc. Am. A \textbf{22}, 1106--1114 (2005).

\bibitem{Schadle2007}
A.~Schädle, L.~Zschiedrich, S.~Burger, R.~Klose, and F.~Schmidt,
  \enquote{Domain decomposition method for Maxwell's equations: scattering off
  periodic structures,} J. Comput. Phys. \textbf{226}, 477 --
  493 (2007).

\bibitem{OWC}
A.~Sommerfeld, \emph{Partial differential equations in physics}
  (Academic Press, New York, 1949).


\bibitem{Zolla:07}
Y.~Ould~Agha, A.~Nicolet, F.~Zolla and S.~Guenneau, \enquote{Leaky modes in twisted
  microstructured optical fibres,} Wave Random Complex \textbf{17}, 559–570 (2007).

\bibitem{Demesy:07}
G.~Dem\'{e}sy, F.~Zolla, A.~Nicolet, M.~Commandr\'{e}, and C.~Fossati,
  \enquote{The finite element method as applied to the diffraction by an
  anisotropic grating,} Opt. Express \textbf{15}, 18089--18102 (2007).

\bibitem{Demesy:09}
G.~Dem\'{e}sy, F.~Zolla, A.~Nicolet, and M.~Commandr\'{e}, \enquote{Versatile
  full-vectorial finite element model for crossed gratings,} Opt. Lett.
  \textbf{34}, 2216--2218 (2009).

\bibitem{Ern}
A.~Ern and J.-L.~Guermond, \enquote{Theory and Practice of Finite Elements,}
 Applied Mathematical Series \textbf{159}, (Springer, New York, 2004).

\bibitem{Li}
L.~Li, \enquote{New formulation of the Fourier modal method for crossed surface-relief gratings,}
 J. Opt. Soc. Am. A \textbf{14}, 2758--2767, (1997).

\end{thebibliography}
\end{document}